# Assessing the livability within the 15-minute city concept based on mobile phone data

Tianqi Wang[1,2], Teemu Jama[1,3,4] & Henrikki Tenkanen[1*]

[1] Department of Built Environment, School of Engineering, Aalto University, Espoo, Finland
[2] Department of Economics, Ecole Polytechnique, Paris, France
[3] Department of Architecture, School of Arts, Aalto University, Espoo, Finland
[4] School of Engineering and Natural Sciences, University of Iceland, Reykjavik, Iceland

* Corresponding author: Henrikki Tenkanen (henrikki.tenkanen@aalto.fi)

**Abstract**

Many cities promote walkability through concepts such as the compact city and 15-minute city to enhance urban livability, yet few methods link spatial walkability features to empirically measured livability and account for temporal dynamics. The method developed for this study uses mobile phone data from the Helsinki Metropolitan Area (Finland) to assess whether commonly used, literature-derived livability indicators (diversity, density, proximity, accessibility) predict observed human activity patterns across different times of day.

We constructed two key dimensions of livability: attractiveness and walkability with quantifiable sub-indicators that were selected based on literature. Our analysis shows that walkability, and even more so the combined livability index, correlates with activity patterns, outperforming the pure attractiveness perspective. However, this relationship is temporally unstable, significantly weakening at night and fluctuating daily. Moreover, based on Geographically Weighted Regression analysis, our results reveal significant spatial variation in the relationship between livability and the intensity of human activities. The findings suggest that traditional urban planning goals, such as functional diversity to enhance walkability, contribute to livability but have a limited impact on the 15-minute city's overall sustainable mobility objectives, necessitating a larger-scale perspective and more functionally profiled approaches for urban development.

**Keywords:** Chrono-urbanism; Urban planning; Mobility data; Entropy weighting method; Nordics

# 1. Introduction

The unprecedented worldwide urbanization since the mid-20th century has resulted in profound changes in the socioeconomic and built environments in the cities of major industrial countries. This has resulted in a high concentration of economic activity in relatively small geographical areas, which we know as cities: urban areas cover approximately 1-3 % of Earth's land area (Liu et al., 2014; Zhao et al., 2021) while being responsible for 70 % of global energy-related $CO_2$ emissions (Wu et al., 2020). Today it is well-known that the growth and densification of the city also brings many problems such as urban pollution, increasing GHG emissions, traffic congestion and health problems (e.g. Berghauser Pont et al., 2021, Fang et al., 2016; Li et al., 2016; Zhan et al., 2018). In addition, urban regions are places of increasing inequality and segregation (e.g. Arvidsson et al., 2023; Kuddus et al., 2020).

As economic activities concentrate, the population follows. This has led architects and urban planners to seek ideas for managing densifying urban development and safeguarding human perspectives in the physical city resulting from blueprints. Especially after the seminal works by much-cited scholars, Edward Glaeser and Richard Florida (Florida, 2014; Glaeser et al., 2001), there has been a noticeable surge in recognition of livability in urban planning and policy. Consequently, concepts such as "15-minute city" and "Compact city" – both of which emphasize the urban features of historical cities, such as walkability and proximity of amenities in enhancing modern human life – have been embraced by planners, scholars and institutions (e.g. Angel et al., 2021; Boyko & Cooper, 2011; EEA & FOEN, 2016; Lwasa et al., 2022). However, some scholars have also raised criticism of these concepts, arguing that the idea of compact cities is outdated (Naumann 2005, Ferreira 2011) and pointing out that quantitatively measuring livability, density as well as the achieved impacts can be challenging (Jama et al., 2024; Page et al., 2024; Zhang et al., 2025).

The dominant discourse, particularly in models like the 15-minute city, is critiqued for its measurements to have an inherent nocturnal bias, concentrating excessively on traditional daytime errands and neglecting the crucial role of the night-time economy in fostering a livable city (Nofre, 2024). Recent deluge of urban and geospatial big data (e.g. anonymized mobile phone data) provides novel ways to explore these patterns at finer spatio-temporal scales. These datasets have been seen to increase the understanding of the connections between various living environment characteristics (produced in urban planning) and human activity patterns, moving beyond static measurements (Bergroth et al., 2022; Sulis et al., 2018). This can also contribute to a better understanding of the intricate relationship between urban dynamics and socioeconomic order (Calafiore et al., 2024; Marin Flores et al., 2025), as well as sustainability of urban life (Dey et al., 2025; Tenkanen et al., 2024; Willberg et al., 2024).

Stemming from these starting points, in this study we aim to build a livability evaluation framework considering attractiveness and walkability, and use mobile-phone data to investigate the relationship between temporal livability and more static urban features facilitated by urban plans. We used the Helsinki Metropolitan Area (Finland) as a case region for the framework development. Helsinki has been a rapidly evolving Nordic urban region. Its urban planning is characterized by a strong public focus on sustainability and a relatively low level of laissez-faire development. This makes the region an excellent case study for examining the impacts of specific built-environment features resulting from urban planning. We aim to answer the following research questions: (1) How to quantify livability within the conceptual paradigm of a 15-minute city, when considering different urban functions and land-use features? (2) Does higher livability in a neighborhood result in higher human activity? (3) How do these patterns vary temporally? (4) Which indicators correlate most with human activities?

By addressing these research questions, we aim to attain a more profound understanding of the relationship between urban environment and human activities under the livability theme.

## 2. Literature review

### 2.1 Urban Livability

Numerous frameworks for assessing livability have also been proposed in academic circles in recent years (Benita et al., 2021; Yu et al., 2022). Early discussions on urban livability can be traced back to urban planning discourses in the mid-20th century, when the rapid urbanization and related challenges necessitated a structured approach to assess city life (Voukkali et al., 2024). Over the decades, livability metrics have evolved from mere environmental and infrastructure considerations to incorporate also economic dimensions (Najafi et al., 2024). However the philosophical background stems from cultural dimensions and social values. Jane Jacobs as one of the most famous advocates of these, and a leading critic of hard modernistic functional values, advanced the theory of urban vitality and underscored the pivotal roles of diverse land use and dense street networks. Her book The Death and Life of Great American Cities (Jacobs, 1961) is often seen as the modern research origin of urban livability. Critiques of modernism from a liveability perspective were also taken up by several architectural theorists. Alongside Jane Jacobs's work, Christopher Alexander's Pattern Language stands out as one of the most influential and most concretely formulated bodies of work in which the architectural components of liveability were systematically examined and articulated to humanise the legacy of modernism (Alexander et al., 1977).

Many studies conducted in the 21st century argue that higher-density cities are more livable (Quastel, 2012; Chowdhury, 2020). Environmental amenities have also been well recognized in much of the literature (Badland et al., 2014; Fu et al., 2019; Xiao et al., 2022), which consists of coverage or spatial access to parks, forests, and water areas (Węziak-Białowolska, 2016; Xiao et al., 2021) as well as environmental pollution with respect to water, solid waste, ambient air and noise in the urban area (Saitluanga, 2013; Xiao et al., 2021). Simultaneously, in the assessment of livability as a phenomena itself, spatial accessibility is invoked as a salient determinant of urban desirability behind the livability (Baobeid et al., 2021; Martino et al., 2021; Zanella et al., 2015). Besides the social and physical dimensions of the urban environment discussed above, it has been shown that residents' individual socioeconomic attributes such as age, education, income, employment, and home ownership also affect their satisfaction with the urban environment (Bérenger et al., 2007; Liang et al., 2020; Wang et al., 2022). Despite the widespread attention urban livability has garnered in recent years, a consensus on its precise definition and key-incredients remains elusive.

Due to the lack of consensus on the definition of livability, contemporary methods for measuring livability vary across academic studies. Some scholars focus on residents' subjective perceptions of their living environment, employing surveys and other tools to construct subjective livability assessment frameworks based on emotions and cognitive evaluations. For example, Gyori et al. (2019) constructed their urban livability assessment framework by evaluating the subjective satisfaction levels of over 400 residents globally regarding their urban living experiences. Other studies emphasize more objectively measurable data and socioeconomic indicators within a given area to develop livability indices. For instance, Zhan et al. (2018) utilized the geographical detector model to investigate the characteristics of urban livability satisfaction obtained through questionnaire surveys and the extent of influence exerted by its determining factors. In recent years, methodologies reliant on socio-economic statistical data and other quantitative geographical information have been more extensively applied. For instance, Cui et al. (2022) employ remote sensing data, air quality indices, and points of interest data to compute the composite urban livability scores for the various districts within Wuhan.

## 2.2 15-Minutes City

The concept of 15-minute city envisions urban environments where essential amenities and daily services are accessible within a 15-minute walk or bike ride from most people's homes. The 15-minute city concept, though implicitly present already in various urban design philosophies from the past, has been recently popularized most notably by the Colombian-French urban planner and researcher, Professor Carlos Moreno. His work, rooted in creating "chrono-urbanism," promotes urban transformations where residents can work, live, and play within close proximity, reducing travel times and enhancing the quality of urban life (Moreno et al., 2021.). Thus, the concept emphasizes proximity, density, diversity, digitalization, connectivity and incorporates a human scale approach to cities (Khavarian-Garmsir et al., 2023). As a planning paradigm it aims to re-establish connections between individuals and their communities for more localized urban living. Hence, the approach presents also a remedy to address the intensifying global issue of traffic congestion (Moreno et al., 2021).

Despite, or perhaps because of the concept's old-fashioned nature, it has gained popularity among urban leaders. Cities such as Paris, Melbourne, and Portland have expressed interest in implementing principles of the 15-minute city as a key strategic goal for planning (Pozoukidou et al., 2021). The 15-minute city concept, as a planning policy, represents an effort to reconcile the decentralizing tendency of traditional modern urban regional planning with the ideal of the compact city, a concept often derived without sufficient consideration of its regional context. The City of Paris, under Mayor Anne Hidalgo, has been at the forefront with its "Ville du quart d'heure" initiative, which aims to decentralize city functions and prioritize pedestrian and cycling infrastructure in these decentralised locations (Dakouré et al., 2023). The combination of service decentralization with locally compact land-use parcels is a repetition of the planning paradigm related to modernistic suburb development in the times of top-down centrally planned economy. Today western societies lack a centrally planned economy, but also the planning practice differs from the past. In the contemporary all-expanding urban landscape, the focus is more on the redevelopment for urban regeneration of existing neighborhoods instead of new greenfield development of suburban units as small satellite garden cities.

Nevertheless, the concept of the 15-minute city has not only become a favorite among politicians but has also received a fair amount of attention from scholars. Some studies suggest that the 15-minute city approach can provide significant societal advantages and offer solutions to contemporary urban challenges (Di Marino et al., 2022). Some studies suggest that the 15-minute city paradigm holds potential to contribute to the reduction of urban carbon emissions while helping to build cities that are more sustainable, efficient, resilient, equitable, and inclusive, hence aligning clearly with the global agenda of Sustainable Development Goal (SDG) 11 (Allam et al., 2022). In the wake of the urban revitalization post the COVID-19 pandemic, the 15-minute city concept has gained further intention (Giles-Corti et al., 2023; UN-HABITAT, 2022). Some scholars see the 15-minute city as one of the most popular urban planning concepts globally as it is believed to foster a more human-centric and livable urban environment (Kissfazekas, 2022; Moreno, 2025).

These widespread promises linked to the concept boil down to the phenomenon of livability and call for advanced measurements to gain empirical evidence behind it. Among numerous studies, researchers have primarily focused on the proximity of service facilities to population based on pedestrian network analysis (Li et al., 2019; Hosford et al., 2022; Jiang et al., 2024). Additionally, studies have analyzed the employment structure under the concept of the 20-minute city (Li et al., 2024) . As the concept relies much on the local service development and the density alone does not ensure them to be emerged, there is a clear need to study urban features behind livability in more detail and in different geographical and temporal scales (Jama et al., 2024). Indeed, in scholarly discourse concerning the concept, the emphasis is not exclusively on the urban conditions within a 15-minute timeframe. Some studies have extended their parameters to explore walkability within 20 or 30 minutes (Capasso Da Silva et al.,2020; Birkenfeld et al., 2023; Willberg et al., 2023). Prior research has also emphasized the necessity of distinguishing between daytime and nighttime periods in the analysis (Nofre, 2024). Some scholars have leveraged methodologies like space syntax into their research to deepen the evaluation and find the more proper parameters for the future "the X-minute city" concepts (Beniamino Murgante et al., 2024). However, because modern urban life cannot be reduced to the framework of static spatial data metrics nor the modelled transportation speed, more diverse data sources must be leveraged to identify and classify the spatial dimensions of livability.

## 2.3 Data-driven approaches to understand urban dynamics

In contemporary urban life, the lifestyles and travel habits of city residents have become increasingly diverse. This shift, coupled with the growing sophistication of urban studies, has rendered novel data sources. Since the beginning of the 21st century, mobile phones have gradually become an essential part of urban residents' daily lives, and mobile phone data has increasingly garnered attention within the academic community (Calabrese et al., 2014). Mobile phone data provides continuous, near real-time information on population movements, enabling a nuanced understanding of urban dynamics. Traditional methods, such as surveys and censuses, offer limited temporal and spatial resolution, whereas mobile phone data captures the spatio-temporal activities of large populations with high granularity (Bergroth et al., 2022).

Currently, mobile phone data has become a pivotal resource in urban studies, offering granular insights into human mobility, urban dynamics, and livability. It has widespread applications in urban studies. One of the most relevant studies to this paper, is a study by Osunkoya and Partanen (2024) who applied mobile phone data in their research as a representative of human activities within Tallinn (Estonia) in order to study the spatial distributions of the 'slow' vitality measures (e.g. age of buildings, street length, mixed land use, urban density, street intersections and socio-economic factors) and mobile phone-based dynamic metrics representing human activities. In this study, however, they do not construct a composite livability index nor conduct geographical regression analysis to understand the relationship between livability features and human activities (as done here), but rather investigate the correlation between socio-economic factors against the human activity patterns (Osunkoya & Partanen, 2024). Zhang et al. (2023) utilized mobile phone data to analyze the behavioral patterns of residents in Shanghai, further examining urban vitality at the street level within Shanghai, while Lin et al. (2021) and Guan et al. (2020) explored the accessibility of urban green spaces in Tokyo and Fuzhou using mobile phone data. Furthermore, mobile phone data has played a crucial role in various studies, such as measurement of the impact of the 15-minute city concept on Barcelona (Graells-Garrido et al. 2021) and depiction of the spatial dynamics of urban vitality in Shenzhen (Tu et al. 2020). Overall, these studies demonstrate how mobile phone data enables a more detailed understanding of urban dynamics for livability and 15 minute city goals.

## 3. Study design

Our research is aimed to build a comprehensive livability assessment framework based on the conducted literature review (Section 2) on relevant livability features, and analyze the livability value, calculated using these features, against recorded human activities (Figure 1). For human activity, we utilize mobile phone data to accurately capture the instantaneous distribution of the population, representing the intensity of human activities in different locations within the study area.

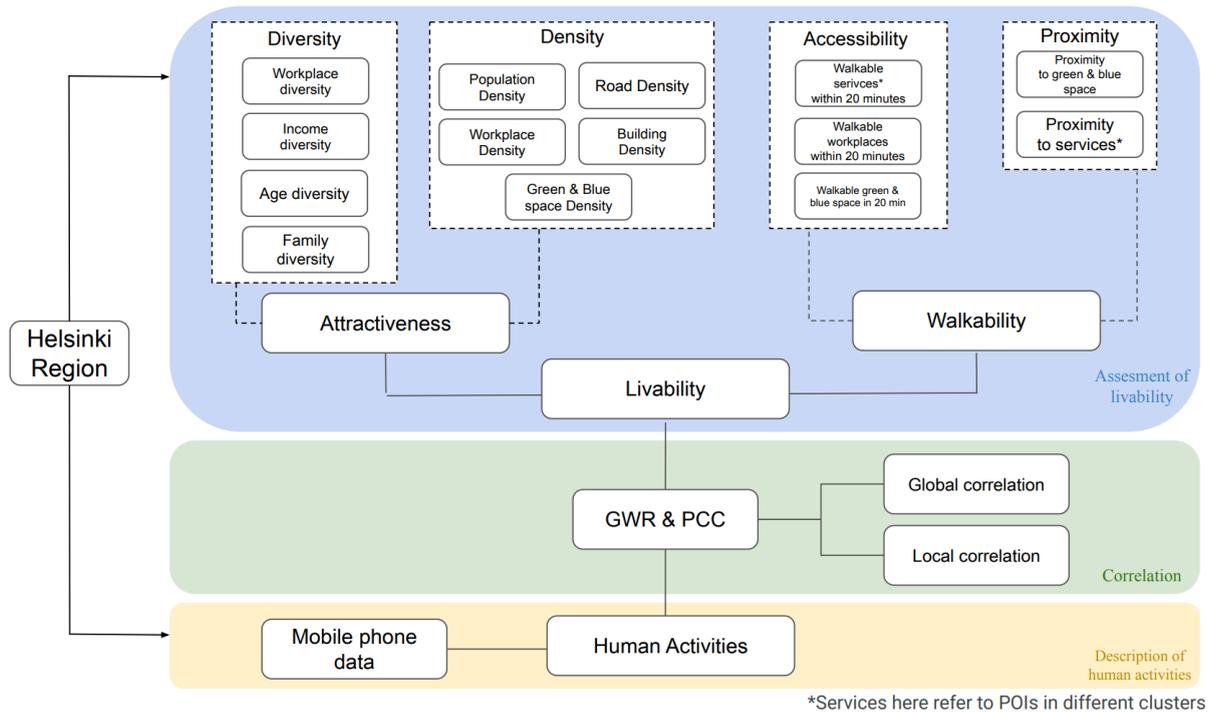

**Figure 1.** Study design and the indicators used to quantify attractiveness, walkability and livability.

For livability, we take into consideration two dimensions: Attractiveness and Walkability. Attractiveness pertains to the qualities and conditions of a grid unit that have the capacity to allure people for purposes such as residence, work, investment, tourism, etc. Walkability refers to the extent to which pedestrian needs are incorporated into urban planning and design, enabling residents in grid units to walk conveniently, safely, and pleasurably to various destinations.

## 4. Data

### 4.1 Mobile phone data

The mobile phone data used in this study present temporally dynamic population distribution in the Helsinki Metropolitan Area (the cities of Helsinki, Espoo, Vantaa and Kauniainen) which is openly available (Bergroth et al., 2022). The data are based on aggregated mobile phone data collected by the biggest mobile network operator in Finland (Elisa Oyj). The dataset is a grid-based dataset at the level of 250 m × 250 m statistical grid cells with hourly proportional distribution of population in each grid cell in regular workdays (Monday to Thursday) and weekends. Given the capacity of mobile phone data to precisely capture instantaneous urban population distributions, here we employ this data to represent human activities within cities. Figure 2 shows an example of how human activity is distributed in our research area according to mobile phone data on a typical Saturday.

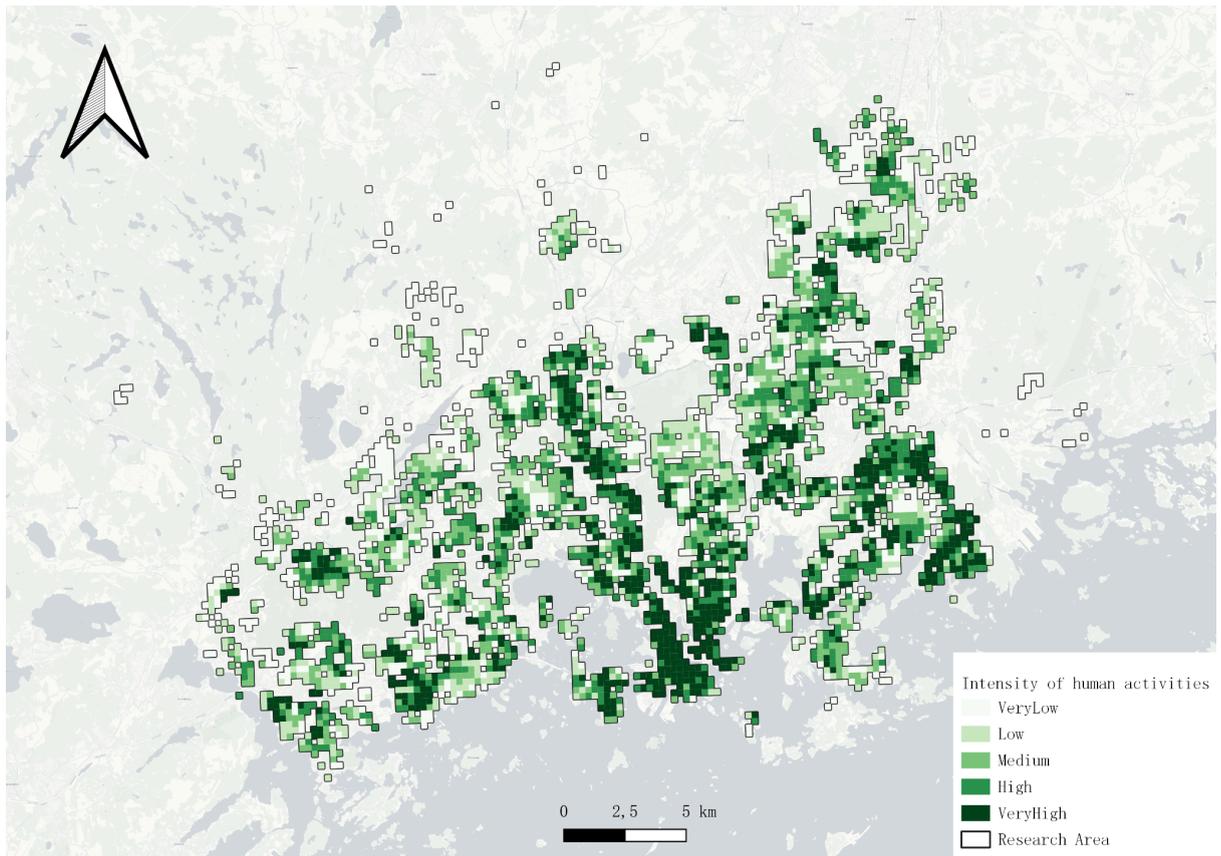

**Figure 2.** The intensity of human activities on Saturday at 11:00 based on mobile phone data.

### 4.2 Socio-economic data

We use Statistics Finland's population register and specifically the grid database for 2018 to obtain socio-economic data for our analysis. The data includes altogether 101 variables in eight groups: Population structure; Educational structure; Inhabitants' disposable monetary income; Size and stage in life of households; Households' disposable monetary income; Buildings and housing; Workplace structure; Main type of activity. Out of these, we use data about workplaces, income, age and family structure to construct the diversity sub-component used in our analysis (see Supplement Table S1 for details). Only grid cells that have over 10 inhabitants and 3 workplaces are used in our analysis (due to privacy constraints).

### 4.3 Points of interest and road network data

We use OpenStreetMap (OSM) to obtain street network data used for accessibility analyses. Moreover, OSM and municipality-maintained Service Maps are used to collect data about Points of Interest including services, such as educational facilities, restaurants, healthcare, leisure and no-turnover services. The locations of grocery stores were collected through publicly available online sources (Pönkänen et al., 2024). Since different clusters of POIs can provide different types of services, POIs in different clusters do not have the same impact on the city. Therefore, we calculate the relevant sub-indices for the different clusters of the POIs separately to make our analysis more robust. A detailed list of POI categories used in the study are listed in the Supplementary materials Table S2.

### 4.4 Other relevant geospatial data

Built and living environment features such as buildings, water bodies, and green spaces were collected from open data sources including public and private organizations, such as Finnish Environment Institute, Helsinki Region Infoshare, and Avoindata.fi. This data was used for calculations related to green & blue areas, and building density.

## 5. Methods

### 5.1 Calculation of indicators

We generated several livability measuring indicators and combined those indicators by weighting them to quantify the level of livability. We give brief descriptions of the indicators in each category below in Table 1. When calculating walkability sub-indices, we considered their typical opening hours and/or times when they are typically used to make our analysis temporally sensitive (13:00 vs 22:00). See Supplement Table S2 for a full list of services considered in the study.

**Table 1**. System of indicators and their descriptions.

| Dimension | Categories | Sub-indices | Description |
|---|---|---|---|
| Attractiveness | Diversity | Workplace diversity | The diversity of workplace in the grid cell |
| | | Income diversity | The diversity of income for the inhabitants in the grid cell |
| | | Age diversity | The diversity of age for the inhabitants in the grid cell |
| | | Family diversity | The diversity of family structure for the households in the grid cell |
| | Density | Residents density | The number of residents per km2 in the grid cell |
| | | Green & blue area density | The percentage of green & blue space in the grid cell |
| | | Workplaces density | The number of workplaces per km2 in the grid cell |
| | | Building density | The percentage of buildings in the grid cell |
| | | Road density | The length of road (m) per km2 in the grid cell |
| Walkability | Accessibility | Walkable regular services within 20 minutes | The number of POIs with daily activities walkable in 20 minutes |
| | | Walkable seldomly used services within 20 minutes | The number of POIs with seldomly activities walkable in 20 minutes |

| | | Walkable workplaces within 20 minutes | The number of workplaces walkable in 20 minutes |
| | | Walkable green & blue space within 20 minutes | The area (m2) of accessible green & blue space within 20 minutes walking |
| | Proximity | Proximity to green & blue space | The distance (m) to the nearest green & blue space |
| | | Proximity to services | The distance (m) to the nearest POIs (separately calculated in clusters) |

For some indicators, we can obtain them directly from existing data or through simple calculations and geospatial analysis techniques (density, accessibility and proximity). However, for diversity, more complex computational methods were required. To compute the subsidiary indices pertinent to diversity, we opted to employ the Gini-Simpson Index, a derivative of the Simpson Index. Edward Hugh Simpson (1949) originally introduced the Simpson Index in 1949, serving as a metric to ascertain the concentration level of individuals categorized into various types.

We divided the workplaces and the inhabitants' income level, age, and family structure into several categories (see Supplement Table S1), which we used to calculate the Simpson's Index λ of each dataset in each grid cells as:

$$\lambda = \sum_{i=1}^{R} p_i^2 \qquad (1)$$

Where $p_i$ is the proportional abundances (the fraction of entities ascribed to each cluster in relation to the entire dataset). R is richness (the cumulative count of distinct types present within the dataset).

Thus, we can then calculate the Gini-Simpson Index as :

$$1 - \lambda = 1 - \sum_{i=1}^{R} p_i^2 \qquad (2)$$

Should the Gini-Simpson Index manifest a high value in a particular grid cell, it indicates a high level of diversity therein. Conversely, a low Gini-Simpson Index value in a grid cell signifies low diversity. Considering the realities of the Helsinki Metropolitan Area, we measure the accessibility of the grid cells by the number of walkable services, workplaces, and green & blue areas within 20 minutes. Based on estimates from previous research, we use 1.29m/s as the standard walking speed in our assessment of walkability (Mohler, B.J., 2007).

All the calculations related to accessibility are processed by QGIS (3.16.15) and Python. We extracted the centroids of the grid cells in our research area and used QGIS to create reachable service areas for each centroid. The convex hull algorithm was used to convert the service areas into buffers, which can tell us the walkable areas of each research grid cell within 20 minutes of walking (see Supplement Figures S1-S3 for visual explanation of the process).

### 5.2 Entropy Weighting Method

We employ the entropy weighting method as the statistical model to evaluate the comprehensive livability in cities. It can give weight to each sub-indices according to its dispersion degree in order to derive a quantifiable livability index for each grid cell.

To construct our livability index, firstly, we construct the sub-indices data matrix:

$$X = (x_{ij})_{m \times n} \tag{3}$$

Where $X_{ij}$ is the value of sub-index $j$ in grid cell $i$. Then, we normalize the values in $X$ within 0 to 100 to ensure comparability across different sub-indices. After that, we calculate the weight of sub-index $j$ in grid cell $i$:

$$P_{ij} = \frac{X_{ij}}{\sum_{i=1}^{n} X_{ij}} \tag{4}$$

Then we calculate the entropy $e_j$ of sub-index $j$ across all grid cells:

$$e_j = - k * \sum_{i=1}^{n} P_{ij} \log(P_{ij}) \tag{5}$$

Where $k = \frac{1}{\ln m}$. Then we calculate the redundancy (importance) of each sub-index:

$$g_j = 1 - e_j \tag{6}$$

A higher $g_j$ value indicates a more important sub-index $j$. Then we compute the weight of each sub-index:

$$W_j = \frac{g_j}{\sum_{j=1}^{m} g_j}, \quad j = 1, 2, ..., m \tag{7}$$

After that, we compute the comprehensive evaluation score for each grid cell:

$$S_i = \sum_{j=1}^{m} W_j * P_{ij}, \quad i = 1, 2, ..., n \tag{8}$$

This yields the final livability index for each grid cell which we use for further analysis. The feature importance of each sub-indicator constructing the livability index are reported in the Supplementary Table S3.

**5.3 Analysis of Correlation: Pearson correlation coefficient & GWR**

In our research, we use Pearson correlation coefficient (PCC) to analyze the global correlation between livability and human activities, and use the Geographically Weighted Regression (GWR) to analyze the local correlation between livability and human activities.

Geographically Weighted Regression (GWR) offers a local approach to examine spatial processes, and takes into account the intrinsic variability across different geographical contexts. It was primarily conceptualized in response to the limitations of traditional regression methods that assume stationary relationships across space (Brunsdon et al., 1996). The GWR model can be expressed as follows:

$$Y_i = X_i \beta_i (u_i, v_i) + \epsilon_i \tag{9}$$

Where $\beta_i (u_i, v_i)$ is the coefficient at the spatial location $(u_i, v_i)$ which is allowed to vary over space. To estimate them in GWR, the weighted least squares estimation is employed:

$$\hat{\beta_i}(u_i, v_i) = (X^T W_i X)^{-1} X^T W_i y \tag{10}$$

Where $W_i$ is a diagonal matrix of weights for the ith observation. The weights are usually determined by a spatial kernel (e.g., Gaussian, bi-square). The most common form of the weighting function is the Gaussian function:

$$\omega_{ij} = e^{\frac{-d_{ij}^2}{2b^2}} \tag{11}$$

Where $\omega_{ij}$ is the weight for the jth observation relative to the ith observation, $d_{ij}$ is the distance between the ith and jth observation, b is the bandwidth, which controls the rate of weight decay with increasing distance.

With this approach, we can estimate specific parameters for every grid cell in our research area to reflect the relationships between independent and dependent variables. Thus, we can assess the local linear relationship between livability and human activities. In our research, we leverage the MGWR application (v2.2) developed by Arizona State University to do the geographically weighted regression.

# 6. Results

## 6.1 Spatial distribution of urban livability in Helsinki Region

In our research, we strategically selected 2 time points to capture distinct periods of livability, ensuring a comprehensive representation of daily rhythms. These time points are: 1) 13:00 on workday, a period typically representative of peak working hours; 2) 22:00 on workday, which reflects residential and resting patterns as people are predominantly at home. Each selected time period provides a snapshot of varying urban dynamics and features, which we posit may have implications for the livability of the area.

a) 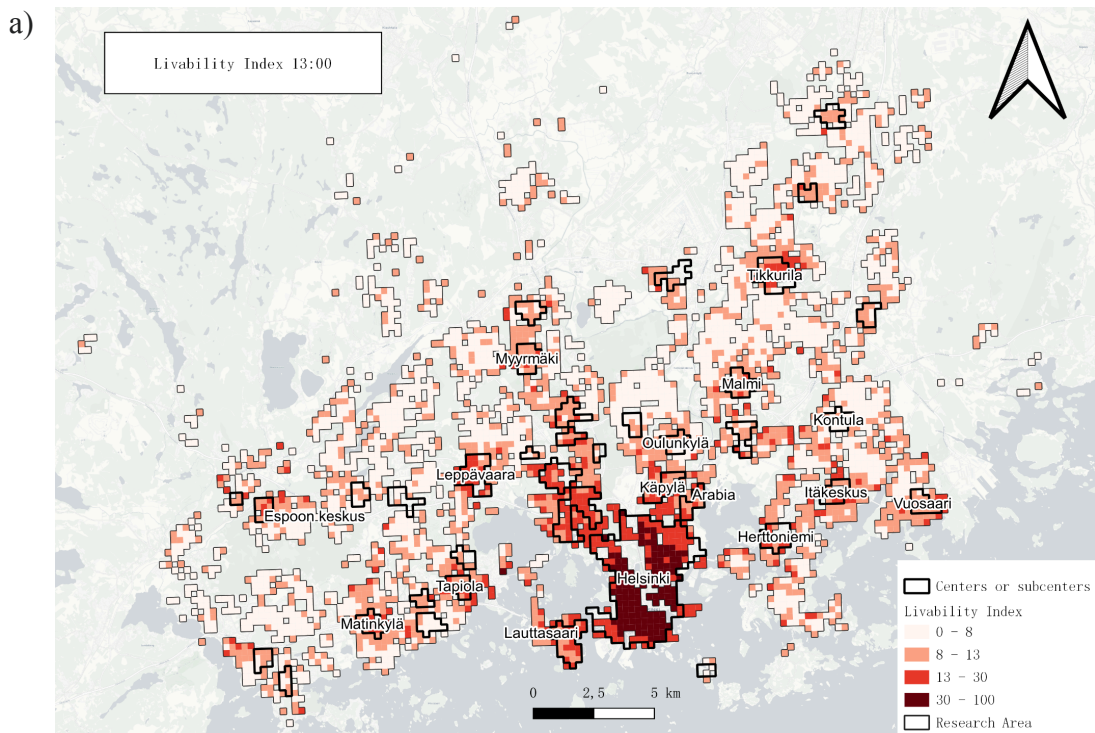

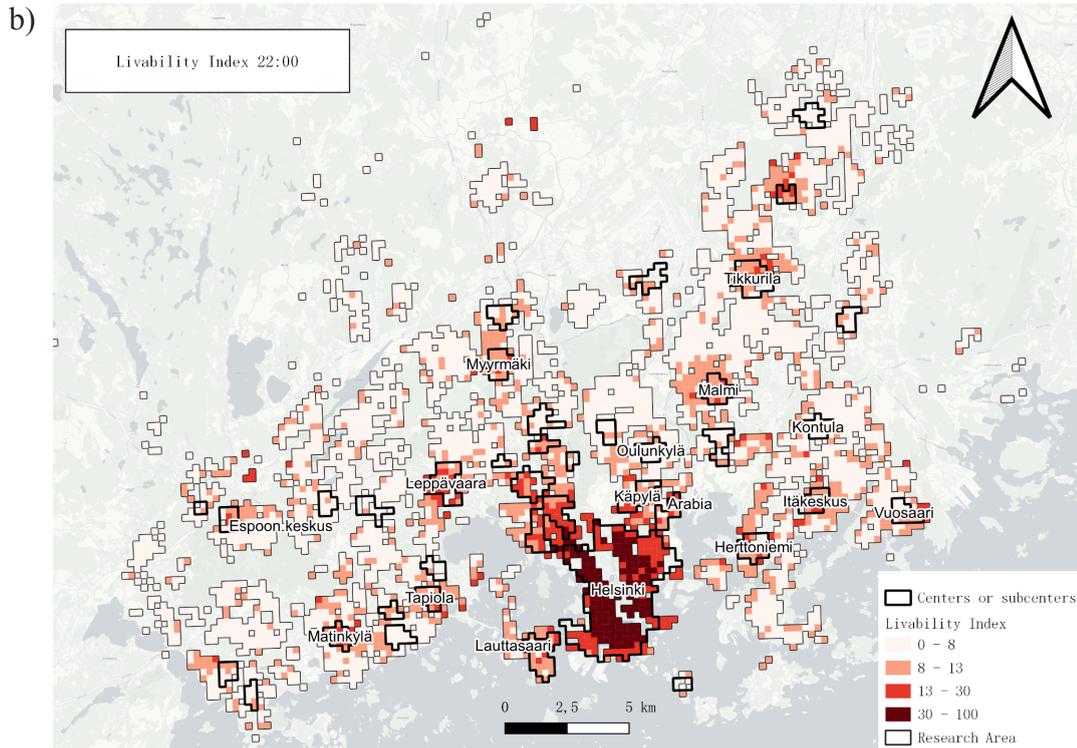

**Figure 3.** Map of livability index in Helsinki region at 13:00 (a) and 22:00 (b).

Figure 3 illustrates the spatial distribution pattern of urban livability at 13:00 and 22:00. Our results reveal that the urban livability index of Helsinki's city center consistently ranks the highest, regardless of whether it is 13:00 or 22:00 on a weekday. The Helsinki city center is the most historically established section of the entire region and unequivocally stands as the economic, commercial, and cultural epicenter of the area. It amasses a significant concentration of businesses, employment opportunities and amenities compared to the rest of the region. In addition, neighborhoods like Tapiola, Arabia, Lauttasaari, Käpylä, Myyrmäki, Espoo Keskus, and Leppävaara also demonstrate higher levels of livability, both at 13:00 and 22:00. Furthermore, it is noteworthy that both the city center and sub-centers consistently exhibit higher livability compared to suburban areas, whether at 13:00 or 22:00. The livability value calculated from the selected features in sub-centers can be attributed to be the result of modernistic master planning history of the Helsinki region: Plots for amenities, jobs and apartment buildings in sub-center locations have been designed using more mixed zoning with denser road networks compared to the rest of the region. This has clearly contributed to the factors behind livability index more than areas outside of the locations even if high residential densities are located also outside of sub-centres. Moreover, it can be observed that areas with higher livability are more spatially dispersed at 13:00 compared to 22:00 indicating the sub-centers have a more limited role in the whole perspective of livable urbanism.

**6.2 The global correlation between livability and the human activities**

To investigate global associations between urban livability and human activities, we employed Pearson correlation coefficient. We investigate the correlations at 13:00 and 22:00 on a workday to capture distinct periods for human activities.

**Table 2.** The result of Pearson correlation coefficient between livability and human activities observed via mobile phone data.

| Index | | Coefficient, r | |
|---|---|---|---|
| | | Time: 13:00 | Time: 22:00 |
| **Attractiveness** | - | **0.48*** | **0.30*** |
| | Diversity | 0.20*** | 0.27*** |
| | Density | 0.46*** | 0.26*** |
| **Walkability** | - | **0.66*** | **0.52*** |
| | Proximity | 0.29*** | 0.23*** |
| | Accessibility | 0.66*** | 0.52*** |
| **Livability** | | **0.68*** | **0.51*** |

Notes=***, ** and * represent 0.001, 0.01 and 0.05 levels of significance

As illustrated in Table 2, there is a notable correlation between urban livability and human activities. The intensity of human activities within each grid cell correlates positively with selected measures of livability. However, the correlation between urban livability and human activities tends to be stronger at 13:00. Concurrently, the dimensions of attractiveness and walkability within the livability value also exhibit higher correlation with the intensity of human activities at this specific time. This suggests that the most pronounced correlation between human activities and urban livability manifests during working hours. This phenomenon can be predominantly attributed to the concentration of individuals in commercial and office districts during working hours, making these areas the epicenter of human activities particularly at these times. Commercial and office districts have typically emerged in the locations with robust regional transportation connectivity. The plethora of amenities and employment opportunities resulting from regional connectivity in these districts further enhances their attractiveness. Consequently, the locational choices of these amenities with the dimensions of attractiveness and walkability measured from the build-up environment in these locations have resulted that human activities and urban livability is clearly higher during this period. However, the impact of regional connectivity, i.e. the fact that some people travel to these locations outside of our study region, can not be deduced from the result.

The correlation between human activities and urban livability is lower at 22:00. This coincides with the period when most individuals have finished their daily work and retreated to their residential areas. During this period, the hot spots of human activities shift to residential districts. Despite population density, the residential areas offer fewer services beyond habitation, hence resulting in a comparatively lower level of attractiveness. This mainly explains why the correlation between livability, attractiveness, and walkability with human activities is weaker at 22:00 compared to 13:00.

## 6.3 The spatio-temporal variation of livability's relationship with human activities

We employed Geographically Weighted Regression (GWR) to investigate the spatial variation in the relationship between the livability indicator (explanatory variable) and human activities (dependent variable). We used adaptive bisquare as the spatial kernel based on K-Nearest Neighbor to specify a neighborhood for a given area with k=50 at 1pm, and k=55 at 10pm. Our geographically weighted regression model demonstrates better performance than the traditional global regression model. At 13:00, the R² of the Geographically Weighted Regression (GWR) model (0.64) was higher than that of the global regression model (0.46), and at 22:00, the R² of the GWR model (0.45) also exceeded that of the global regression model (0.26).

Figure 4 illustrates the spatial distribution of the GWR-based local coefficients of livability at different time periods. These local coefficients inform about how the relationship between a predictor and the response varies across space. We can clearly observe that the spatial distribution patterns of the relationship between livability and human activity differ significantly between 13:00 and 22:00, and overall there is significant geographical variation in this relationship indicating spatial non-stationarity.

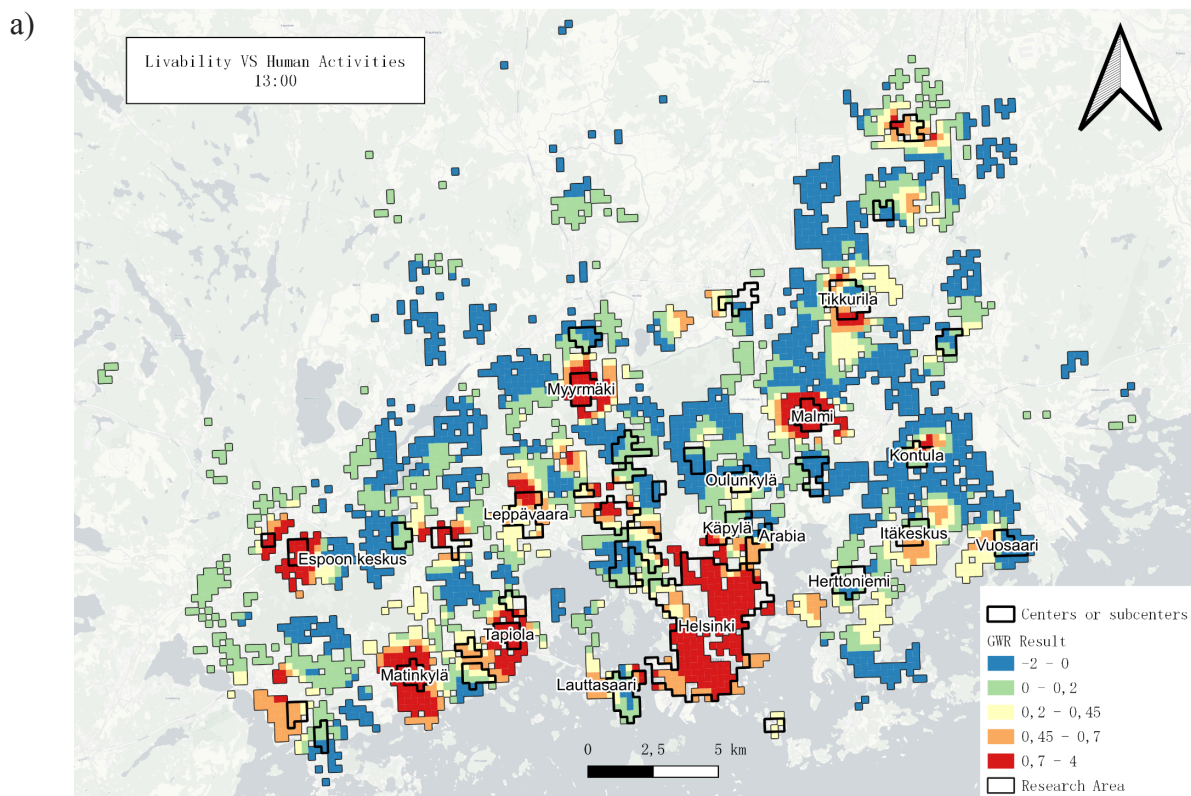

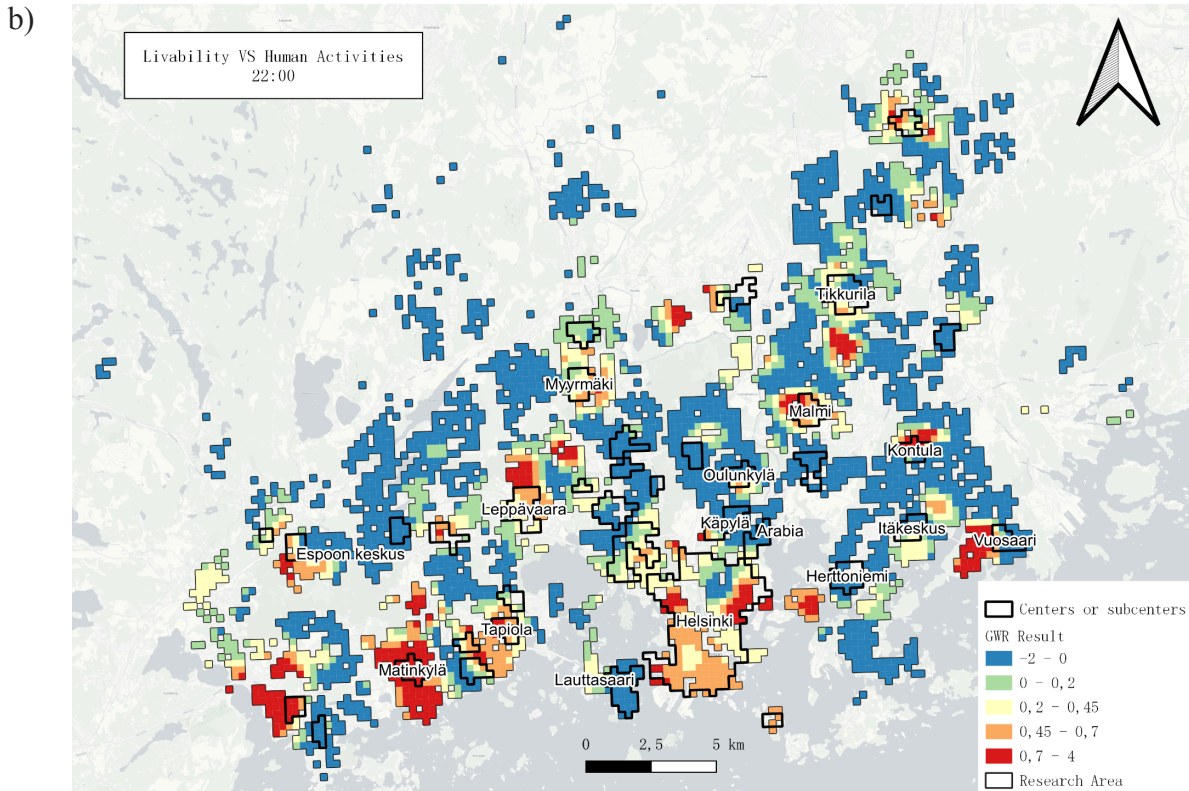

**Figure 4.** Spatial distribution of the GWR local coefficients of Livability at (a) 13:00; (b) 22:00.

The results reveal that areas with negative and lower local coefficients are more extensive at 22:00. It is apparent that at 13:00, there is a strong positive relationship between human activity and livability especially in the city center of Helsinki. However, this relationship noticeably weakens at 22:00. This same phenomenon is also observed in Tapiola, Myyrmäki, Tikkurila, Espoon keskus, and Leppävaara. However, in Vuosaari, we observe a significant increase in positive relationship between human activity and livability indicators at 22:00 compared to 13:00. The primary reason for this is most likely the difference in human activity patterns between daytime and nighttime on weekdays. This daytime concentration of human activity in these highly livable sub-centers results in a strong relationship between livability and human presence. However, due to urban structure, most workers live in suburban areas where livability is relatively lower leading to a decrease in weighted human activity, even though their livability remains unchanged, resulting in a weakened local coefficient. This has an especially clear impact in the city region which has a strong functional zoning tradition in planning and less mixed-use neighborhoods. However, Vuosaari, being a sub-center with a stronger residential character, exhibits a stronger positive relationship between variables at 22:00, which can also be explained by this pattern.

In addition, there are certain similarities in the spatial distribution patterns of the local GWR coefficient between human activity and livability during the daytime and nighttime on weekdays. Firstly, whether during the daytime or nighttime on weekdays, areas where human activity is negatively associated with livability are mostly located in the suburbs, while areas with a strong relationship between human activity and livability are predominantly in city centers and sub-centers, even though this relationship tends to weaken at night. Secondly, during both the daytime and nighttime, there is a strong positive relationship between human activity and livability in Malmi and Matinkylä. This may indicate that these two areas also have a relatively good balance between employment and residential living even though it can not be deduced whether the population and the human activity are locally linked.

**6.4 The impacts of different indices on human activities**

To demonstrate the strength of the correlation of each subcomponent on human activity intensity in our livability assessment framework, we computed the Pearson correlation coefficient for each subcomponent within our livability framework in relation to the intensity of human activity at 13:00 and 22:00. Table 3 presents the results of the Pearson correlation coefficients.

**Table 3.** Correlation coefficients between different indicators and the human activity data at 13:00 and 22:00.

| Indicator | Pearson r - 13:00 | Pearson r - 22:00 | Component |
|---|---|---|---|
| Age Diversity | 0.061*** | 0.146*** | Attractiveness - Diversity |
| Family Diversity | 0.354*** | 0.306*** | |
| Income Diversity | -0.118*** | -0.058*** | |
| Job Diversity | 0.269*** | 0.264*** | |
| Building density | 0.567*** | 0.396*** | Attractiveness - Density |
| Blue-Green Area density | -0.201*** | -0.208*** | |
| Road Density | 0.185*** | 0.253*** | |
| Density Residents | 0.545*** | 0.527*** | |
| Density Job | 0.410*** | 0.200*** | |
| Prox Bank | -0.421*** | - | Walkability - Proximity |
| Prox Edu | -0.382*** | - | |
| Prox Food | -0.482*** | -0.331*** | |
| Prox Grocery | -0.432*** | -0.327*** | |
| Prox Green | 0.156*** | 0.155*** | |
| Prox Leisure | -0.356*** | -0.197*** | |
| Prox NoTurnover | -0.433*** | - | |
| Prox Healthcare | -0.489*** | -0.189*** | |
| Regularly 20 | 0.481*** | 0.312*** | Walkability - Accessibility |
| Seldomly 20 | 0.509*** | 0.221*** | |
| Green 20 | 0.155*** | 0.046** | |
| Job 20 | 0.509*** | 0.321*** | |

Notes=***, ** and * represent 0.001, 0.01 and 0.05 levels of significance

The results reveal that at 13:00, the most positively correlated subcomponent with the intensity of human activity is the walkability related subcomponent. This might suggest that walkability features may have a positive impact on the intensity of human activity, supporting the benefits of the 15-minute city planning concept. But as we can not deduce the indirect contribution of the regional connectivity, nor the impact of modernistic zoning (lack of facilities for services out-side sub-centers), as explained earlier, we can not conclude the connection. Both the number of services and workplaces accessible within a 20-minute walk exhibit a strong positive correlation with human activity intensity. However, at 22:00, there is a noticeable decline in the correlation levels for both service walkability and job walkability. Their correlations with the intensity of human activity decreases significantly during this time. Again, this aligns with our expectations, as daytime human activity on weekdays is primarily related to work and commercial activities, which explains the strong positive correlation with sub-indices associated with these locations. In the evening, human activity is predominantly residential, leading to the significant decline in the positive correlations mentioned earlier. This impact of modernistic zoning can be also seen when looking at it from a green and blues -space access perspective: Both 13:00 and 22:00, the area of green and blue space accessible within a 20-minute walk demonstrates a positive correlation with the intensity of human activity. This indicates a high level of planned greenery in the sub-urban Helsinki Region, where all grid units have walkable access to green spaces.

## 7. Discussion

In this study, we aimed to shed light on the features linked to the concept of the 15 minute city from the perspective of Jane Jacobs' theories. We created quantifiable measures of urban functional diversity, structural compactness, and walkability to investigate the geographical patterns of livability in the Helsinki region across space and time.

### 7.1. Livability as a dynamic phenomenon

Our study, using Helsinki Metropolitan Areas as case, shows that the selected livability features are mostly positively associated with the intensity of human activity which is mainly pronounced during daytime. This is in line with the findings in the similar recent livability assessments (Osunkoya & Partanen, 2024). Our results show that the walkability-related livability sub-indices have the strongest and most systematic positive correlation related to the intensity of human activity, while attractiveness (density, diversity) related sub-indices have a less profound relationship. During night-time, however, the relationship between livability and human activity intensity in urban centers and sub-centers generally diminishes. Consequently, our findings demonstrate that livability is a phenomenon that fluctuates as a function of time and place, which also is inline with previous studies (Farber et al., 2014; Järv et al., 2018; Tenkanen et al., 2016; Widener & Shannon, 2014). Some scholars have pointed out that urban planning often prioritizes the "daytime city," leading to a significant disconnect between the goals driven by urban plans and the reality of how cities function around the clock (Nofre, 2024). Planning for temporal dynamics is vital for social equity, as millions of people around the world work in logistics, healthcare, essential services, and the informal economy around the clock, also during night-time hours (Nofre, 2024; Nofre & Garcia-Ruiz, 2023). The traffic that the compact city and 15-minute city concepts seek to reduce is a constant, round-the-clock phenomenon, driven by factors that extend far beyond conventional metrics of density or diversity.

The comparison of day-time and night-time to each other also reveals the need to profile livability values for different lifestyles. Liveable daily life is not the same for all the residents but the urban facilities which support one's livability and needs varies. Moreover, the type of needs might also vary in time for the same residents, making it dynamic while the features of the land-use in particular location is very static. This variance creates the need for urban planners to take this into consideration and facilitate a more diverse set of livability features for neighbourhoods.

**7.2 Geographical variation of livability**

Our results show that during the day, the relationship between livability and human activity is strongest in the more central locations of the city (sub-centers) which have high regional connectivity as well as the highest share of human activity. During the night-time the relationship weakens in these central areas (even in areas where amenities remain open at night time) and shifts towards suburbs as the impact of regional traffic and workplaces diminishes. The findings suggest that the Helsinki region, which still remains relatively unicentric, exhibits behaviour primarily centered around day-time activities. Conversely, results from global cities such as London would likely show a more balanced distribution between day-time and night-time activities, particularly within their central business districts.

Overall, even if derived from a globally small city region, our results are aligned with the findings of Liang et al. (2022) regarding the linkage of walkability and attractiveness to urban vitality. We did not find any location which would have had high human activity without any livability features. However, positive correlation between green area features and human activity in locations which lack the typical urban features facilitated by urban design (i.e., red pixels clearly outside of service center areas indicated with black boundaries), raises the importance of the features located inside them to be considered as a land-use type for livability. To integrate a more diverse set of features also in the well reachable sub-center locations would support sustainable livability more effectively than just a set of traditional urban amenities like cafes, restaurants or shopping that focus on consumption.

**7.3 Limitations**

There are naturally some limitations in our study. For instance, due to limited access to relevant data, we did not incorporate indicators related to urban safety into our analysis, which are proved to have impact on the livability (Buys et al., 2012; Tao, Wong, & Hui, 2014; Fleming et al., 2016; Fu et al., 2019). Our livability assessment framework also predominantly centers on objectively identifiable urban resources and services and does not consider the specific needs of certain population groups which vary based on various socio-economic aspects (Willberg et al., 2024). Furthermore, it is important to understand that the legitimated spaces for the facilities which contribute the livability in a given area is ultimately a result of urban plan-making following the strategies for aimed service networks facilitated with regional transportation investments. Hence, they tend to follow the logic of placement for accordingly centralized services where regional accessibility is the driver. In addition, the use of traditional "Jacobsian" indicators for walkability and attractiveness, often end up promoting existing city centers as urban design paragons for livability while neglecting the fact that in central locations both the population sum and the diversity of services are high because of the regional impact on them. Meanwhile in the suburban region the services do not follow the planned density and hence the densification alone does not support the 15-minute ideal (Jama et al. 2024). These regional feedback loops behind land-use plan details (for walkability) complicates the interpretation of the results toward causality hypotheses and clearly calls for a transdisciplinary and multi-scale perspective with accordingly structured data.

Considering these limitations, the proposed urban livability assessment framework and the demonstrated research methodology can be extended to bring more diverse quantitative insights for plan-making decisions and the used metrics.

**7.4 Avenues for future research**

Future research would benefit from using a similar approach to study livability in relation to observed human activities but incorporating information on citizens' home area and daily visited places. That would give more nuanced insights, as the impact of regional connectivity could be deduced and the insights about people's real activity spaces could be incorporated to the analysis (Abbiasov et al., 2024). Activity spaces can be further used to study how city regions land-use plans and traffic investments support 15-minute city end-goal for more decentralized, equal and balanced land-use, which is prerequirement for less carbon intensive lifestyle patterns (Raudsepp et al., 2024).

Our study showcases that mobile phone based data provides useful information about the dynamic intensity of urban human activities across different time periods that can enrich the livability assessments for the ideals like 15-minute city. As the cities are highly dynamic by nature, we recommend the broader application of mobile phone data in future urban studies to better understand how different areas of the city are used by its varied citizens. Furthermore, future studies could consider incorporating methodologies such as sentiment analysis to objectively capture residents' subjective perceptions (Liu et al., 2020), enhancing the precision of evaluations.

Murgante, B., Valluzzi, R., & Annunziata, A. (2023). Developing a 15-minute city: Evaluating urban quality using configurational analysis. The case study of Terni and Matera, Italy. *Applied Geography*, *162*, 103171.

Najafi, E., Hosseinali, F., Najafi, M. M., & Sharifi, A. (2024). A GIS-based Evaluation of Urban Livability using Factor Analysis and a Combination of Environmental and Socio-economic Indicators. *Journal of Geovisualization and Spatial Analysis*, *8*(2).

Neuman, M. (2005). "The Compact City Fallacy." *Journal of Planning Education and Research* 25, 1: 11–26.

Nofre, J. (2024). The need to design a nocturnal 15-min city. *Urban Geography*, 45(7), 1267–1277.

Nofre, J., & Garcia-Ruiz, M. (2023). Nightlife Studies: Past, Present and Future. *Journal of Electronic Dance Music Culture* 15(1).

Osunkoya, K. M., & Partanen, J. (2024). Enhancing urban vitality: integrating traditional metrics with big data and socio-economic insights. *Journal of Spatial Information Science*, *29*, 43–68.

Page, M., Joutsiniemi, A., Vaattovaara, M., Jama, T. & Rönnberg O. (2024). Density as an Indicator of Sustainable Urban Development: Insights from Helsinki? *European Planning Studies* 32(10): 2182–202.

Pönkänen, M., Tenkanen, H., & Mladenović, M. (2024). Spatial accessibility and transport inequity in Finland: Open source models and perspectives from planning practice. *Computers Environment and Urban Systems*, 116, 102218.

Pozoukidou, G., & Chatziyiannaki, Z. (2021). 15-Minute City: Decomposing the new urban Planning eutopia. *Sustainability*, *13*(2), 928.

Quastel, N., Moos, M., & Lynch, N. (2012). Sustainability-As-Density and the Return of the Social: The Case of Vancouver, British Columbia. *Urban Geography*, 33(7), 1055–1084.

Raudsepp J, Czepkiewicz M, Heinonen J, et al. (2024) Travel footprints in the nordics. *Environmental Research Communications* 6(9): 095002.

Saitluanga, B. L. (2013). Spatial pattern of urban livability in Himalayan region: a case of Aizawl City, India. *Social Indicators Research*, *117*(2), 541–559.

Simpson, E. H. (1949). Measurement of diversity. *Nature*, *163*(4148), 688.

Sulis, P., Manley, E., Zhong, C. & Batty, M. (2018). Using Mobility Data as Proxy for Measuring Urban Vitality. *Journal of Spatial Information Science* 16: 137–62.

Tenkanen, H., Chau, N., & Dey, S. (2024). A data-driven approach for estimating travel-related carbon emissions with high spatial and temporal granularity. *AGILE GIScience Series*, *5*, 1–4.

## Acknowledgements

ChatGPT (GPT-o4) was used for language improvements during the writing of the paper to clarify and correct grammar.

## Declaration of interests

The authors declare that they have no known competing financial interests or personal relationships that could have appeared to influence the work reported in this paper.

## Data Availability

The data that support the findings of this study is mostly open source (referenced in the text) except the Statistics Finland's population grid data at 250 meter resolution which cannot be shared openly.

## Supplementary material

**Table S1**. Data attributes used to construct the sub-indices of the study.

| Sub-indices | Dataset involved | Attributes |
|---|---|---|
| Workplace diversity | The workplaces provided within the grid cells | Agriculture, forestry and fishing; Mining and quarrying; Manufacturing; Electricity, gas, steam and air conditioning supply; Water supply, sewerage, waste management and remediation activities; Construction; Wholesale and retail trade, repair of motor vehicles and motorcycles; Transportation and storage; Accommodation and food service activities; Information and communication; Financial and insurance activities; Real estate activities; Professional, scientific and technical activities; Administrative and support service activities; Public administration and defense, compulsory social security; Education; Human health and social work activities; Arts, entertainment and recreation; Other service activities; Activities of households as employers, undifferentiated goods- and services-producing activities of households for own use; Activities of |

| | | extraterritorial organizations and bodies |
|---|---|---|
| Income diversity | The income level of inhabitants within the grid cells | Lowest income; Middle income; Highest income |
| Age diversity | The age of inhabitants within the grid cells | 0-2 years old; 3-6 years old; 7-12 years old; 13-15 years old; 16-17 years old; 18-19 years old; 20-24 years old; 25-29 years old; 30-34 years old; 35-39 years old; 40-44 years old; 45-49 years old; 50-54 years old; 55-59 years old; 60-64 years old; 65-69 years old; 70-74 years old; 75-79 years old; 80-84 years old; Over 84 years old |
| Family diversity | The family structure of inhabitants within the grid cells | Young single persons (aged 19 to 35); Young couples without children (aged 19 to 35); Households with at least one child under 18; Middle-aged households (aged 36 to 64); Pensioner households (over 65) |

**Table S2.** OpenStreetMap, Palvelukartta, and Grocery attributes and the classification used in the study.

| Classification | Sub-classification | Attributes |
|---|---|---|
| Regular services | Restaurant | Bar, Biergarten, Cafe, Fast food, Food court, Pub, Restaurant |
| | Education | Kindergartens, Elementary, Middle school, High school, Universities |
| | Grocery | S-market, K-market, Lidl, M-market |
| Seldom used services | Bank | Bank |
| | Healthcare | University hospitals, Acute psychiatry, Pharmacy, Aid services, City hospitals, City psychiatric hospital, X-rays, Laboratory, Domestic hospitals, Maternity hospitals, Social care work for adults, Doctor's reception, Health stations, Health center on-call service and emergency outpatient clinic for adults, Diabetes nurse reception, Nurse's reception, Social work for adults, Cervical cancer screenings, Emergency care outpatient wards, Child health clinic, Hospital outpatient clinics, Palliative care |

|  | Leisure | Art center, Casino, Cinema, Concert hall, Music rehearsal place, Piano, Spa, Swimming, Theater |
|  | No turnover | Libraries, Place of worship, Events venue |

**Table S3.** The feature importance of each indicator

| Index | Weighting - 13:00 | Weighting - 22:00 | Component |
|---|---|---|---|
| Age Diversity | 0.000524 | 0.000453 | Attractiveness - Diversity |
| Family Diversity | 0.001183 | 0.001024 | |
| Income Diversity | 0.001394 | 0.001206 | |
| Job Diversity | 0.020344 | 0.017613 | |
| Building density | 0.020598 | 0.017833 | Attractiveness - Density |
| Blue-Green Area density | 0.115395 | 0.099904 | |
| Road Density | 0.023498 | 0.020344 | |
| Density Residents | 0.062475 | 0.054088 | |
| Density Job | 0.255519 | 0.221218 | |
| Prox Bank | 0.001915 | - | Walkability - Proximity |
| Prox Edu | 0.000398 | - | |
| Prox Food | 0.000551 | 0.000765 | |
| Prox Grocery | 0.000447 | 0.000878 | |
| Prox Green | 0.001974 | 0.001709 | |
| Prox Leisure | 0.004291 | 0.002372 | |
| Prox NoTurnover | 0.001122 | - | |
| Prox Healthcare | 0.000926 | 0.002811 | |
| Regularly 20 | 0.150484 | 0.163867 | Walkability - Accessibility |
| Seldomly 20 | 0.164388 | 0.244505 | |
| Green 20 | 0.026889 | 0.023280 | |
| Job 20 | 0.145688 | 0.126131 | |

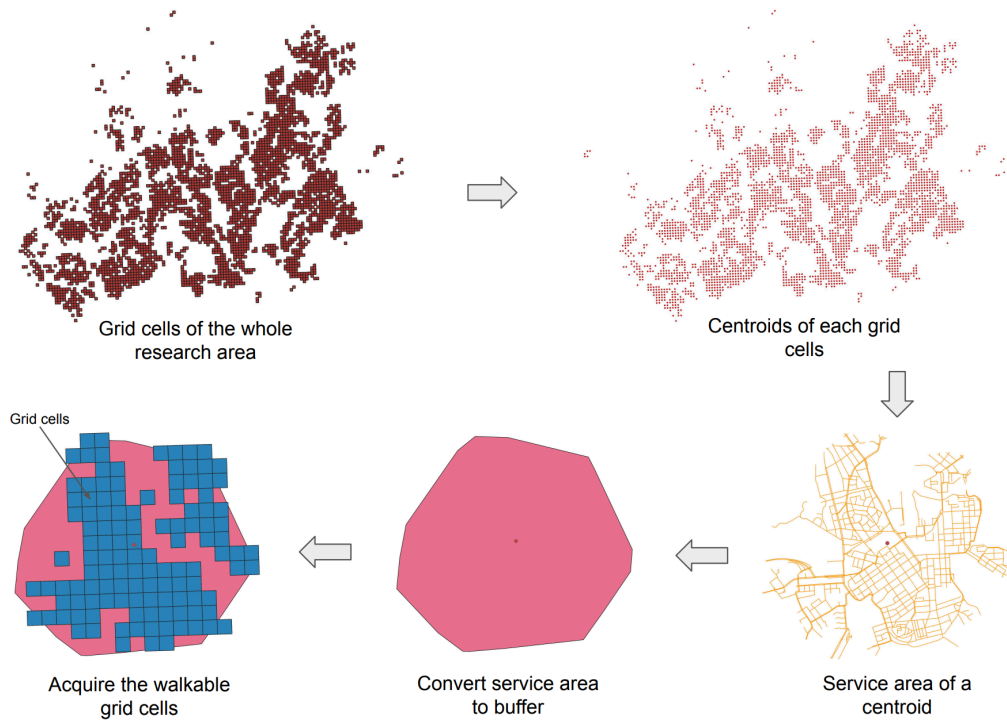

**Figure S1.** Calculation process of walkable workplaces

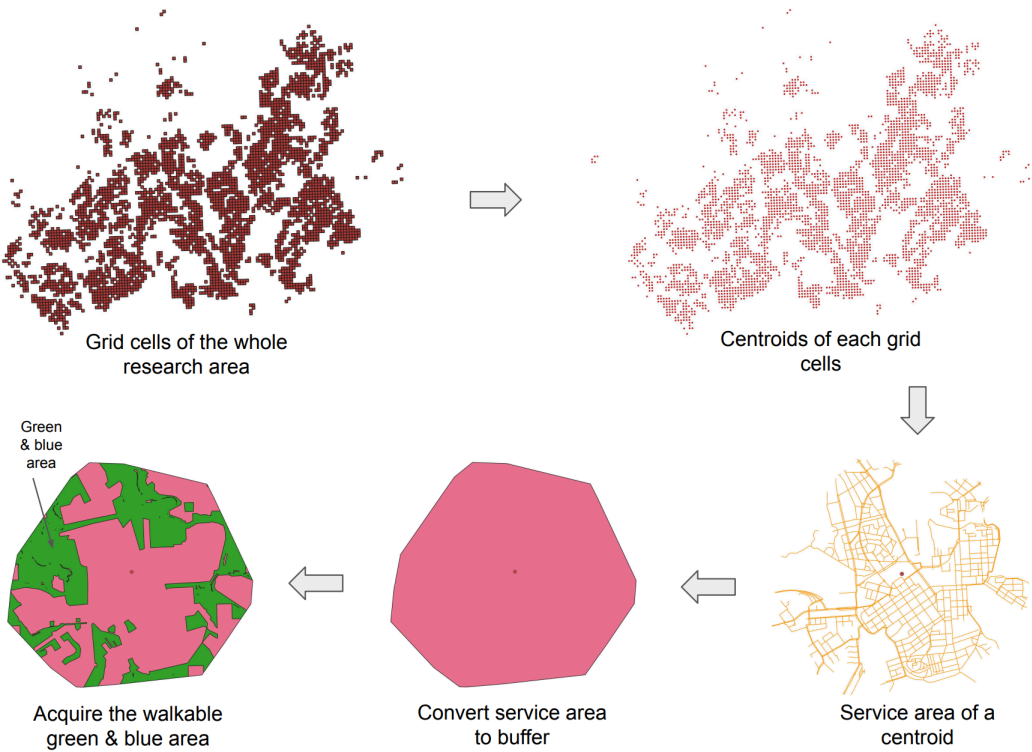

**Figure S2.** Calculation process of walkable green & blue areas

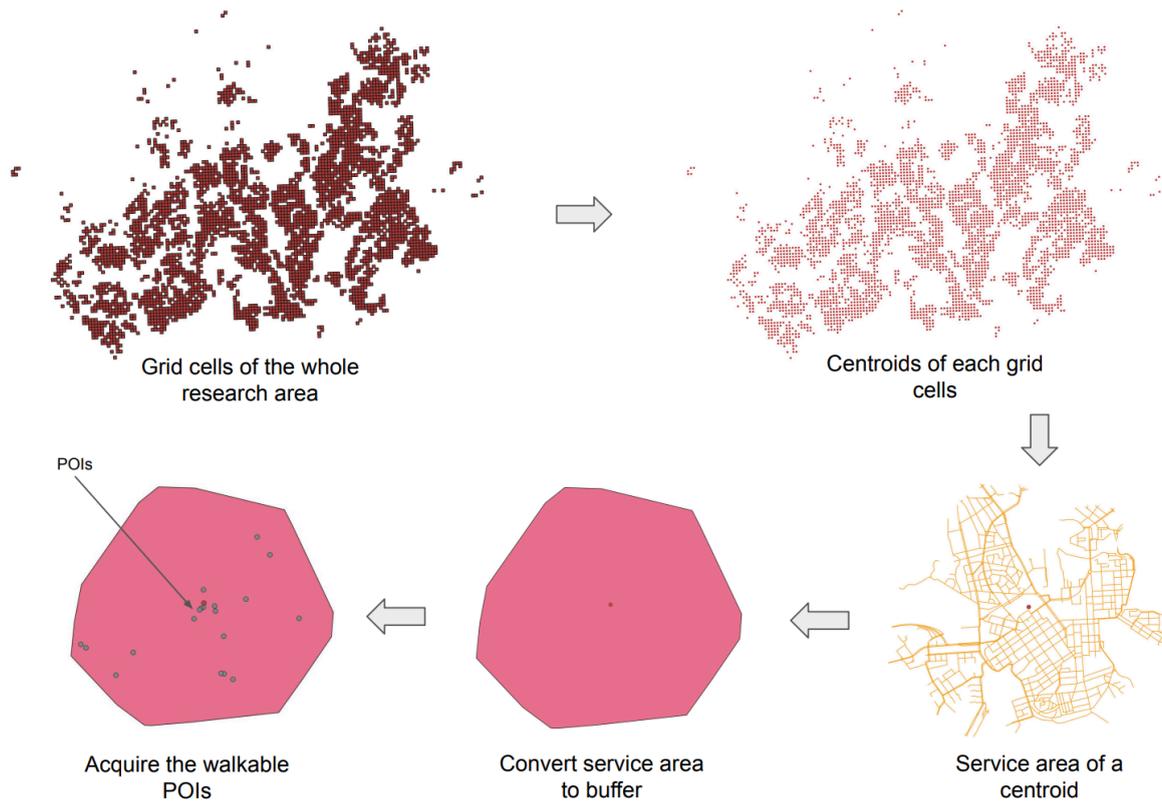

**Figure S3.** Calculation process of walkable services